\documentclass[prd,amsfonts,twocolumn,nofootinbib,showpacs,dvipdfmx]{revtex4}
\RequirePackage{graphicx}
\usepackage{enumerate}
\usepackage{bm}
\usepackage{braket}
\usepackage{slashed}
\usepackage{comment}
\usepackage{color}
\usepackage{ulem}
\begin{document}
\title{The exact WKB for cosmological particle production}

\author{Seishi Enomoto}
\affiliation{School of Physics, Sun Yat-sen University, Guangzhou 510275, China}
\author{Tomohiro Matsuda}
\affiliation{Laboratory of Physics, Saitama Institute of Technology,
Fukaya, Saitama 369-0293, Japan}
\begin{abstract}
The Bogoliubov transformation in cosmological particle production
can be explained by the Stokes phenomena of the corresponding ordinary
differential equation.
The calculation becomes very simple as far as the solution is
 described by a special function.
Otherwise, the calculation requires more tactics, where 
the Exact WKB (EWKB) may be a powerful tool.
Using the EWKB, we discuss cosmological particle production focusing on
 the effect of more general interaction and classical scattering. 
The classical scattering appears when the corresponding scattering
 problem of the Schr\"odinger equation develops classical turning points
 on the trajectory. 
The higher process of fermionic preheating is also discussed using the
 Landau-Zener model.
\end{abstract}

\maketitle
\section{Introduction}
Particles may be formed from the vacuum when fundamental parameters such
as mass or interaction coefficients become time-dependent, and
there are many reasons for which the fundamental parameters could change during
cosmological evolution.
Among them, particle production caused by inflaton oscillation is
known to be very important for solving the problem of reheating the
universe after inflation\cite{Traschen:1990sw, Kofman:1997yn,Kofman:2004yc}. 

The motion of inflaton field is a damped oscillation.
However, at least near the center of the oscillation, where particle
production is likely to take place, linear approximation with respect to
$t$ can be made. 
Then, one can write the inflaton motion as $\phi(t)=vt$, which
drastically simplifies the problem.
Typically, the mass of a scalar field (e.g, $\chi$) is supposed to be
given by 
\begin{eqnarray}
m^2_\chi(t)&=&m_0^2 +g^2_2 \phi(t)^2,
\end{eqnarray}
where $\phi(t)$ is the oscillating inflaton field.
If we consider the Lagrangian given by
\begin{eqnarray}
{\cal L}_\chi&=&\frac{1}{2}\partial_\mu \chi\partial^\mu \chi
 -\frac{1}{2}m_0^2\chi^2 -\frac{1}{2}g_2^2 \phi(t)^2\chi^2,
\end{eqnarray}
the equation of motion is given by 
\begin{eqnarray}
\frac{d^2 \chi}{dt^2}+\left[k^2+m^2_\chi(t)\right]\chi=0.
\end{eqnarray}
If one replaces $\phi(t)$ with $\phi(t)\simeq vt$, the above equation is
equivalent to the Schr\"odinger equation of the scattering problem by the
``inverted quadratic potential'' given by
\begin{eqnarray}
V(t)&=&-\left(g_2^2v^2\right)t^2,
\end{eqnarray}
where the corresponding ``energy'' is 
\begin{eqnarray}
E&=&k^2+m_0^2.
\end{eqnarray}
Note that $E>V$ is always true in this case.
Therefore, there is no classical turning point in the scattering
problem.
The particle formation (preheating) with $E<V$ is discussed in
Ref.\cite{Felder:2000hj} and succeeding papers.
Sometimes the phenomenon is called tachyonic preheating, since $E<V$
is realized when $m_\chi^2<0$. 

There are a variety of methods for finding the wave functions of the
general one-dimensional scattering problem of quantum mechanics.
For the inverted quadratic potential, one can find the exact solution 
(i.e, the Weber function, or the parabolic cylinder functions).
Alternatively, one can calculate the scattering coefficients
approximating the potential by a series of steps\cite{Berry:1972na}.
This gives the integral of the coefficients.
Analytic continuation of the WKB expansion and related
topics have a long history\cite{Pokrovskii:1961}. 
For instance, it has been applied to pair production in vacuum by an alternating
field\cite{Brezin:1970xf}.
The Schwinger mechanism\cite{Schwinger:1951nm}, which is named after
Schwinger who first derived the exponential formula for the pair
production, can be analyzed using the complex WKB, and is still an
active research target\cite{Shakeri:2019mnt,  
Kitamoto:2020tjm}. 
For cosmological preheating, a similar calculation has been given by Chung in Ref.
\cite{Chung:1998bt}.
In Ref.\cite{Enomoto:2013mla,Enomoto:2014hza}, the method has been used
to calculate the effect of higher-dimensional interaction during
preheating.
In this paper, we consider the Exact WKB analysis (EWKB)\cite{Voros:1983,Delabaere:1993} for
the particle formation with some exotic interactions.
The EWKB is a powerful tool for calculating the connection
formula, especially useful when the entire solution cannot be transformed into a special
function. 
Using the EWKB, we analyze the effects of higher-dimensional interaction and the
classical scattering of bosons and fermions.
The reader can compare these results with our earlier
calculation\cite{Enomoto:2013mla,Enomoto:2014hza} reviewed in 
Appendix \ref{sec:steepest_descent_method}.
In our next paper, we are going to examine asymmetric particle formation
(kinematic baryogenesis) using the results obtained in this
paper\cite{matsuda_to_appear}.

\subsection{Cosmological preheating as the scattering problem of the
  Schr\"odinger equation with inverted quadratic potential for $E>0$}
Before using the EWKB method to cosmological preheating, 
let us remember how particle production has been treated in the
typical cosmological preheating scenario.
Here, we ignore the expansion factor of the universe for simplicity,
since our model considers almost instant particle production and also
the factor can be included by redefining the parameters. 
Typically, the WKB expansion is used to find
\begin{eqnarray}
\label{eq-WKB}
\chi_k(t)&=& \frac{\alpha_k(t)}{\sqrt{2\omega_k}}e^{-i \int^t \omega dt}
+\frac{\beta_k(t)}{\sqrt{2\omega}}e^{+i \int^t \omega dt},
\end{eqnarray}
where
\begin{eqnarray}
\omega_k(t)&\equiv&k^2+m_\chi^2(t).
\end{eqnarray}
We take $\alpha_k=1, \beta_k=0$ for the initial vacuum state.
The distribution of the particle in the final state is 
\begin{eqnarray}
n_\chi(k)=|\beta_k|^2,
\end{eqnarray}
which can be found by solving the scattering problem of the
corresponding Schr\"odinger equation.
For the above model (i.e, scattering by the inverted quadratic
potential), the following Weber equation
\begin{eqnarray}
y''(z)+\left(\nu+\frac{1}{2}-\frac{1}{4}z^2\right)y(z)=0
\end{eqnarray}
has the solution $D_\nu(z), D_{-\nu-1}(iz)$.\footnote{Note that the
following relation 
\begin{eqnarray}
D_\nu(z)&=&e^{i\nu\pi}D_\nu(-z)+\frac{\sqrt{2\pi}}{\Gamma(-\nu)}
 e^{i(\nu+1)\pi/2}D_{-\nu-1}(-iz)\nonumber
\end{eqnarray}
tells that both $D_\nu(-z)$ and $D_{-\nu-1}(iz)$ are also the solutions
of the equation, although they are not linearly independent.}
More specifically, one can define
\begin{eqnarray}
z&\equiv& ie^{i\pi/4}\sqrt{2g_2v}t
\end{eqnarray}
in the original field equation to find 
\begin{eqnarray}
\frac{d^2 \chi}{dz^2}+\left[\nu+\frac{1}{2} -\frac{1}{4}z^2\right]\chi=0.
\end{eqnarray}
Here we defined
\begin{eqnarray}
\nu=\frac{k^2+m_0^2}{2g_2v}i-\frac{1}{2},
\end{eqnarray}
and for later use we define 
\begin{eqnarray}
\kappa&\equiv& \frac{k^2+m_0^2}{2g_2v}
\end{eqnarray}
and 
\begin{eqnarray}
\nu=i\kappa -\frac{1}{2}.
\end{eqnarray}
Here, $\kappa$ is an important parameter, which is later used to
 estimate the particle production.
The asymptotic forms are given by
\begin{eqnarray}
1.&& |\mathrm{arg} z|<\frac{3\pi}{4}\nonumber\\
D_\nu(z)&\rightarrow& e^{-z^2/4}z^\nu,\\
2.&&-\frac{5}{4}\pi<\mathrm{arg} z<-\frac{\pi}{4}\nonumber\\
D_\nu(z)&\rightarrow&
 e^{-\frac{z^2}{4}}z^\nu-\frac{\sqrt{2\pi}}{\Gamma(-\nu)}e^{-i\nu\pi+\frac{z^2}{4}}z^{-\nu-1}, \\
3.&&\frac{\pi}{4}<\mathrm{arg} z<\frac{5\pi}{4}\nonumber\\
D_\nu(z)&\rightarrow&
 e^{-\frac{z^2}{4}}z^\nu-\frac{\sqrt{2\pi}}{\Gamma(-\nu)}e^{i\nu\pi+\frac{z^2}{4}}z^{-\nu-1}.
\end{eqnarray}
Since $z\equiv i e^{i\pi/4}\sqrt{2g_2v}t$ is used here, $t<0$
gives $\frac{5\pi}{4}<\mathrm{arg}z<\frac{9\pi}{4}$, which corresponds
to the region 1.
Also, $t\rightarrow +\infty$ corresponds to the region 3.
Therefore, we find for $t\rightarrow -\infty$,
\begin{eqnarray}
e^{-\frac{z^2}{4}}&=&e^{-i\frac{g_2v}{2}t^2}\\
z^\nu&=&e^{(i\kappa-\frac{1}{2})\log z}\nonumber\\
&=&e^{(i\kappa-\frac{1}{2})\left(\log(\sqrt{2g_2 v} |t|)+i\frac{3\pi}{4}\right)},
\end{eqnarray}
which gives ($t=-|t|=e^{\pi i}|t|$ is used here)
\begin{eqnarray}
D_\nu(z)&\simeq&
e^{-i\frac{g_2v}{2}t^2}e^{(i\kappa-\frac{1}{2})\left(\log(\sqrt{2g_2 v}
|t|)-i\frac{\pi}{4}\right)},\\
D_{-\nu-1}(iz)&\simeq& e^{+i\frac{g_2v}{2}t^2}e^{(-i\kappa-\frac{1}{2})\left(\log(\sqrt{2g_2 v}
						 |t|)+i\frac{\pi}{4}\right)}.
\end{eqnarray}
Note that the above solutions in the limit $t\rightarrow -\infty$ are
giving the $\pm$ WKB solutions of Eq.(\ref{eq-WKB}).
Therefore, we define 
\begin{eqnarray}
\chi_- &\rightarrow&D_{\nu}(z)\\
\chi_+ &\rightarrow&D_{-\nu-1}(iz).
\end{eqnarray}

On the other hand, in the $t\rightarrow +\infty$ limit we find
\begin{eqnarray}
e^{-\frac{z^2}{4}}&=&e^{-i\frac{g_2v}{2}t^2}\\
z^\nu&=&e^{(i\kappa-\frac{1}{2})\log z}\nonumber\\
&=&e^{(i\kappa-\frac{1}{2})\left(\log(\sqrt{2g_2 v} t)  +i\frac{3\pi}{4}\right)},
\end{eqnarray}
which gives in this limit,
\begin{eqnarray}
D_\nu(z)&\simeq& e^{-i\frac{g_2v}{2}t^2}e^{(i\kappa+\frac{1}{2})
\left(\log(\sqrt{2g_2 v}t)+i\frac{3\pi}{4}\right)}\nonumber\\
&&+i\frac{\sqrt{2\pi}}{\Gamma(-\nu)}e^{i\frac{g_2v}{2}t^2}
e^{-\kappa \pi}e^{(-i\kappa-\frac{1}{2})\left(\log(\sqrt{2g_2 v}t)
+i\frac{3\pi}{4}\right)}.\nonumber\\
\end{eqnarray}
Immediately, one will find that in the $t=+\infty$ limit the asymptotic
form of the exact solution $D_\nu(z)$ is the mixture of the
$\pm$ WKB solutions, which gives the connection formula.
In this case, the connection formula gives the Bogoliubov
transformation of the WKB solutions.
In the calculation of the connection formula, we use
\begin{eqnarray}
\Gamma(z)\Gamma(1-z)&=&\frac{\pi}{\sin \pi z}\\
\Gamma(\bar{z})&=&\overline{\Gamma(z)}\\
1+\nu&=&1+\left(i\kappa-\frac{1}{2}\right)=-\overline{\nu}
\end{eqnarray}
for the calculation of $\Gamma(-\nu)=\Gamma(-i\kappa+\frac{1}{2})$.
This gives 
\begin{eqnarray}
|\Gamma(-\nu)|^2&=&\frac{\pi}{\sin \pi (-\nu)}\nonumber\\
&=&\frac{2\pi i}{e^{-i\pi\nu}-e^{i\pi\nu}}\nonumber\\
&=&\frac{2\pi}{e^{\pi\kappa}+e^{-\pi\kappa}}\\
\Gamma(-\nu)&=&\frac{\sqrt{2\pi}e^{-\pi\kappa/2}}{\sqrt{1+e^{-2\pi\kappa}}}
e^{\mathrm{arg}\Gamma(-\nu)}.
\end{eqnarray}
Finally, one obtains the connection formula given by
\begin{eqnarray}
\left(
\begin{array}{c}
\alpha_k^R\\
\beta_k^R
\end{array}
\right)
&=&
\left(
\begin{array}{cc}
\sqrt{1+e^{-2\pi \kappa}}e^{i\theta_1} & ie^{-\pi\kappa+i\theta_2}\\
-ie^{-\pi\kappa-i\theta_2} &\sqrt{1+e^{-2\pi \kappa}}e^{-i\theta_1} 
\end{array}
\right)
\left(
\begin{array}{c}
\alpha_k^L\\
\beta_k^L
\end{array}
\right),\nonumber\\
\end{eqnarray}
where L and R are for $t\rightarrow -\infty$ and $t\rightarrow +\infty$, respectively.
Here, all the phase parameters are included in $\theta_{1,2}(k)$.
Viewing the result as the solution of the scattering problem, the
reflection and the penetration amplitudes are 
\begin{eqnarray}
|R_k|&=&\frac{e^{-\pi \kappa}}{\sqrt{1+e^{-2\pi \kappa}}}\nonumber\\
|T_k|&=&\frac{1}{\sqrt{1+e^{-2\pi \kappa}}}.
\end{eqnarray}

The above calculation can be obtained by approximating the potential by
a series of steps.
This alternative approach gives 
\begin{eqnarray}
 \dot{\alpha}_k &=& \beta_k \frac{\dot{\omega}_k}{2\omega_k} e^{+2i\int_{-\infty}^t dt' \omega_k}\\
 \dot{\beta}_k &=& \alpha_k \frac{\dot{\omega}_k}{2\omega_k} e^{-2i\int_{-\infty}^t dt' \omega_k}, \label{eq:EOM_of_beta}
\end{eqnarray}
which can be integrated to give the connection
formula\cite{Berry:1972na,Chung:1998bt,Enomoto:2013mla,Enomoto:2014hza}.
Note that the analytic continuation is possible for the above integration.

In the above (simple) scenario, there is no classical reflection point
(turning point) on the real axis, which means that classically the
reflection is not allowed in the scattering problem. 
Particle production becomes significant when $\kappa <1$, where the
quantum scattering process becomes significant.

On the other hand, it would be natural to think about the effect of
classical particle production caused by a classical turning point.
Is the significant amplification of the particle production possible in
this case?
This is the primary question in this paper.
The difference from the tachyonic preheating scenario will be discussed
clearly in the text.

\subsection{Preheating as scattering with the quadratic potential $E<0$}
\begin{figure}[ht]
\centering
\includegraphics[width=0.9\columnwidth]{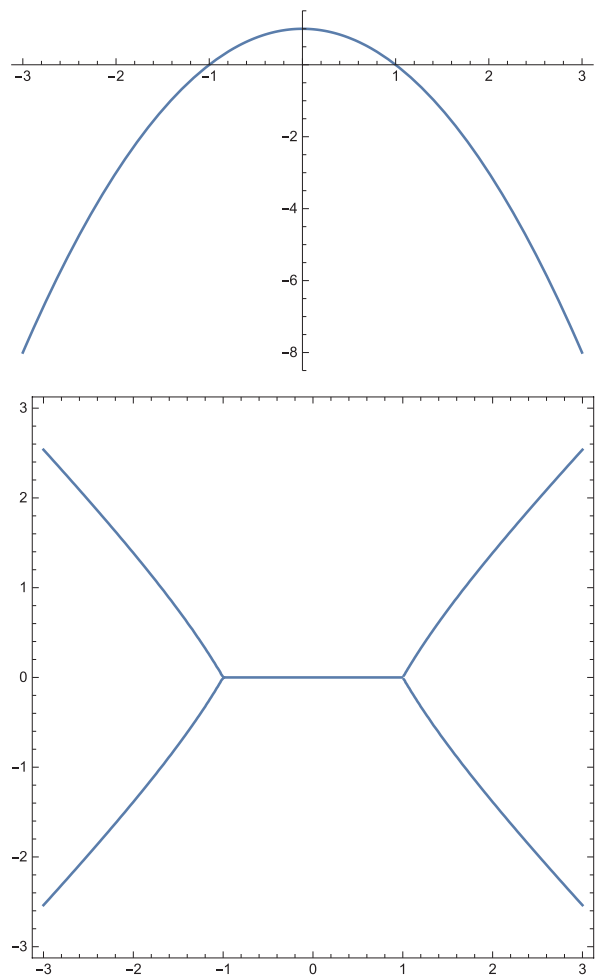}
 \caption{The potential and the Stokes lines are shown for $E<0$.
In this picture, $\chi$ is tachyonic within $-1<x<1$.
The classical turning points are appearing at $x=\pm 1$.}
\label{fig_tach-pot}
\end{figure}

For later use, we analyze particle creation when $m^2_\chi(t)<0$ is
realized temporary.
The typical potential and the Stokes lines of the model are given in
Fig.\ref{fig_tach-pot}.

The situation is similar to the tachyonic preheating scenario after
hybrid inflation\cite{Felder:2000hj}.
However, if one identifies $\chi$ as the waterfall field of the hybrid
inflation model, the field is already tachyonic from the beginning of
the oscillation.\footnote{There could be some exceptional models of
hybrid inflation\cite{Lazarides:1995vr,Matsuda:2009yj,Matsuda:2003ke},
 which changes the conditions discussed in this section. Here, we will
focus on the standard scenario of hybrid inflation.}
In this case, one is not calculating the scattering by the classical
turning point, since the starting point of the process is already
between the two turning points(see Fig.\ref{fig_tach-pot}) and the later
process occurs within the tachyonic region.
Note that the end of hybrid inflation is defined by $m_\chi(\phi_c)^2=0$, which
corresponds to the turning point of Fig.\ref{fig_tach-pot}.
(We are neglecting $k$ just for simplicity.)
This is not the situation we are considering in this paper.
We are considering the situation when significant particle
production is caused by the scattering by the classical turning point,
in which $m_\chi^2<0$ is 
temporal.\footnote{Since the particle creation may cause significant
trapping of the inflaton\cite{Enomoto:2013mla, Kofman:2004yc,
Matsuda:2007tr}, which may trap the inflaton within the tachyonic
region after the particle production, we are focusing on the first
particle production after inflation.}
Therefore, although the potential and the interaction might be similar to
the hybrid inflation model, we are not considering hybrid inflation.
To avoid hybrid inflation, we assume $V(\chi)\ll V_I$, where $V_I$
denotes the potential during inflation, and $V(\chi)$ is the potential
introduced for $\chi$.

In our case, to realize the classical turning points, the original mass
$m_0^2$ has a negative sign. 
Then, we have $E=k^2-m_0^2$ for the corresponding Schr\"odinger equation,
which realizes $E<0$ during the scattering as far as $k<k_0\equiv m_0$.
Our assumption $V(\chi)\ll V_I$ suggests that classical turning points
are expected to appear near the center of the oscillation.
In this case, the nonadiabatic condition is not important for the
particle production, since the scattering is ``classical''.
(Note that ``classical'' is used here for the classical scattering
process in the scattering problem.)
Using the exact solution (i.e, the Weber function), one can easily find
the solution to the scattering problem.
The difference from the previous solution appears in $\nu$, whose
imaginary part becomes negative in this case.
This flip of the sign exchanges the classical and the quantum contributions.
Let us see what happens specifically.
The equation of motion is 
\begin{eqnarray}
\frac{d^2 \chi}{dt^2}+\left[k^2-m_0^2+g^2_2v^2 t^2 \right]\chi&=&0,
\end{eqnarray}
where the corresponding  Schr\"odinger equation has 
\begin{eqnarray}
E_k&=&k^2-m_0^2\\
V(t)&=&-\left(g_2^2v^2\right)t^2.
\end{eqnarray}
For $k<k_0$, classical turning points appear at $t_*=\pm
\frac{\sqrt{-k^2+m_0^2}}{g_2v}$ on the real axis of $t$, and $\chi$ is
tachyonic within
$-\frac{\sqrt{-k^2+m_0^2}}{g_2v}<t<\frac{\sqrt{-k^2+m_0^2}}{g_2v}$.
Since the difference appears in $\nu$, we redefine
\begin{eqnarray}
\hat{\nu}&\equiv&\frac{k^2-m_0^2}{2g_2v}i-\frac{1}{2}
\end{eqnarray}
and 
\begin{eqnarray}
\hat{\kappa}&\equiv& \frac{-k^2+m_0^2}{2g_2v}>0,
\end{eqnarray}
where 
\begin{eqnarray}
\hat{\nu}=-i\hat{\kappa} -\frac{1}{2}.
\end{eqnarray}
To make $\hat{\kappa}>0$, the sign of $\hat{\kappa}$ is taken to be
opposite to $\kappa$.
The Weber function gives 
\begin{eqnarray}
\left(
\begin{array}{c}
\alpha_k^R\\
\beta_k^R
\end{array}
\right)
&=&
\left(
\begin{array}{cc}
\sqrt{1+e^{2\pi \hat{\kappa}}}e^{i\hat{\theta}_1} & ie^{\pi\hat{\kappa}+i\hat{\theta}_2}\\
-ie^{\pi\hat{\kappa}-i\hat{\theta}_2} &\sqrt{1+e^{2\pi \hat{\kappa}}}e^{-i\hat{\theta}_1} 
\end{array}
\right)
\left(
\begin{array}{c}
\alpha_k^L\\
\beta_k^L
\end{array}
\right),\nonumber\\
\end{eqnarray}
which leads to 
\begin{eqnarray}
|R_k|&=&\frac{e^{\pi \hat{\kappa}}}{\sqrt{1+e^{2\pi \hat{\kappa}}}}\nonumber\\
&=&\frac{1}{\sqrt{1+e^{-2\pi \hat{\kappa}}}}\nonumber\\
|T_k|&=&\frac{1}{\sqrt{1+e^{2\pi \hat{\kappa}}}}\nonumber\\
&=&\frac{e^{-\pi \hat{\kappa}}}{\sqrt{1+e^{-2\pi \hat{\kappa}}}},\nonumber\\
\end{eqnarray}
as expected.

Now our questions are:
\begin{enumerate}
\item Under what conditions can we expect meaningful amplification of 
      particle production by the classical turning points?
\item Do the conditions necessary for the amplification meet the
      cosmological requirements?
\end{enumerate}
Before discussing the details, to avoid confusion, we have to say that
such amplification is unlikely to occur as far as the quadratic potential (i.e,
the conventional $g^2\phi^2\chi^2$ interaction) is considered for the particle
production.
We are not claiming that the result is new for the community.
Perhaps, this is the reason why the classical scattering
has not been discussed for the cosmological preheating scenarios.
In this paper, we extend the analysis to include higher terms to find a
reasonable amplification by the classical scattering.
The amplification appearing for the higher interaction is new.

Let us first specify the meaning of ``amplification'' discussed above.
A typical preheating scenario considers significant particle production within the Fermi sphere
$k<k_*\equiv \sqrt{2 g_2 v}$.
Therefore, one can estimate $n\sim k_*^3$ for the scenario.
If one expects ``amplification'' of the particle production, it is
reasonable to examine whether $k_0>k_*$ is realized in the model or not.
We find $k_0>k_*$ gives
\begin{eqnarray}
\label{eq-const1}
m_0^2&>&2g_2 v,
\end{eqnarray}
which leads to
\begin{eqnarray}
g_2&<& 10^{-8}
\left(\frac{m_0}{10^{12}\mathrm{GeV}}\right)^2
\left(\frac{10^{16}\mathrm{GeV}}{\sqrt{v}}\right)^2.
\end{eqnarray}
This alone is a simple fine-tuning of the parameters.

If one wants to avoid tachyonic $\chi$ before the onset of
the inflaton oscillation, 
one has to consider 
another condition $m_0 < g_2 \phi_0$, where $\phi_0$ is the initial
amplitude of the oscillation.
Therefore, one has
\begin{eqnarray}
\label{eq-const2}
g_2^2 \phi_0^2>&m_0^2&>2g_2 v.
\end{eqnarray}
The condition gives the lower bound given by
\begin{eqnarray}
g_2&>& 10^{-4}
\left(\frac{10^{18}\mathrm{GeV}}{\phi_0}\right)^2
\left(\frac{\sqrt{v}}{10^{16}\mathrm{GeV}}\right)^2.
\end{eqnarray}
Apparently, these conditions are contradicting.
Therefore, at least in this model, the negative mass term does not cause 
``amplification by the classical scattering''.
 
Similar particle production could be possible when bubbles (cosmological
domain walls) collide.\footnote{For domain walls made from
superpotential, the cosmological domain wall problem is not serious
as far as the fine-tuning of the vacuum energy is realized by a constant
term in the superpotential\cite{Matsuda:1998ms}}
Unfortunately again, we could not find significant amplification in such a scenario.

Therefore, to find significant ``amplification by the classical
scattering'',  we are going to extend the scenario and include
higher interaction.

\section{The Exact WKB for preheating}

In this section, we describe the calculation based on the EWKB analysis, which gives a
simple method for calculating the connection formula.
Theoretically, the basic connection formulae are obtained by adding up
all the WKB expansions and taking the Borel summation.
The merit of using the EWKB is that once the Stokes lines are given, the
connection formula is manifest for the Borel-summed WKB solutions.
Note that the normalization problem is typical in the ordinary
WKB method, which can be avoided by putting consistency
conditions\cite{Berry:1972na}. 
The normalization problem in the EWKB has been discussed and solved in
Ref.\cite{Voros:1983,Silverstone:2008, Aoki:2009} for typical MTP
(Merged pair of simple Turning Points).
In this case, the factor has to be introduced on both sides of each MTP
structure.

To avoid confusion, we note here that for the simple example discussed here
(i.e, scattering with an inverted quadratic potential), the exact solution
is already known as the Weber function and there is no obvious merit
for considering alternative formulation.
We are introducing the EWKB here for later convenience.

\subsection{Connection formulae by EWKB}
Typical EWKB uses $\eta\equiv \hbar^{-1}\gg 1$, instead of using the
Planck constant.
Following Ref.\cite{Virtual:2015HKT}, our starting point is the
``Schr\"odinger equation'' in quantum mechanics given by 
\begin{eqnarray}
\left[-\frac{d^2}{dx^2}+\eta^2 Q(x)
\right]\psi(x,\eta)&=&0,
\end{eqnarray}
where 
\begin{eqnarray}
Q(x)&=&V(x)-E
\end{eqnarray}
for the potential $V$ and the energy $E$.

\begin{figure}[ht]
\centering
\includegraphics[width=0.9\columnwidth]{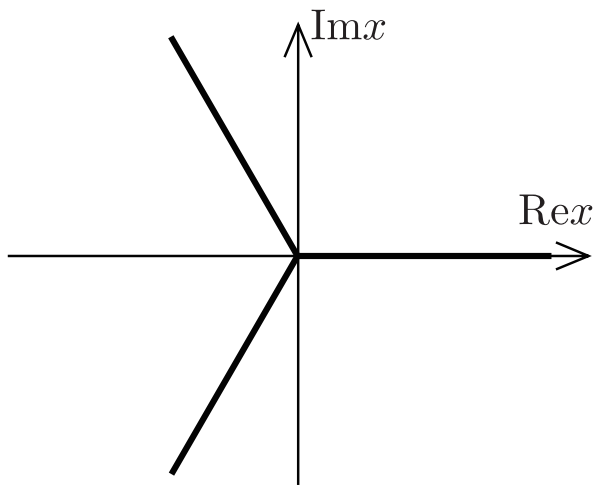}
 \caption{The Stokes lines for $Q=x$.}
\label{fig_airy}
\end{figure}

If the solution $\psi$ is written as $\psi(x,\eta)=e^{R(x,\eta)}$,
we have 
\begin{eqnarray}
\psi&=&e^{\int^x_{x_0}S(x,\eta)dx}
\end{eqnarray}
for $S(x,\eta)\equiv \partial R/\partial x$.
For $S$, we have 
\begin{eqnarray}
-\left(S^2 +\frac{\partial S}{\partial x}\right)+\eta^2 Q&=&0.
\end{eqnarray}
If one expands $S$ as $S(x,\eta)=\sum_{n=-1}^{n=\infty}\eta^{-n} S_{n}$,
one will find
\begin{eqnarray}
S=\eta S_{-1}(x)+ S_0(x)+\eta^{-1}S_1(x)+...,
\end{eqnarray}
which leads
\begin{eqnarray}
S_{-1}^2&=&Q\\
2S_{-1}S_j&=&-\left[\sum_{k+l=j-1,k\ge 0,l\ge 0}S_kS_l + \frac{d
	       S_{j-1}}{dx}\right]\\
&&(j\ge 0).\nonumber
\end{eqnarray}
Using the relation between the odd and the even series, one will have 
\begin{eqnarray}
\psi&=&\frac{1}{\sqrt{S_{odd}}}e^{\int^x_{x_0}S_{odd}dx}\\
&&S_{odd}\equiv\sum_{j\ge 0}\eta^{1-2j}S_{2j-1}.
\end{eqnarray}
Depending on the sign of the first $S_{-1}=\pm \sqrt{Q(x)}$, there are
two solutions $\psi_\pm$, which are given by
\begin{eqnarray}
\psi_{\pm}&=&\frac{1}{\sqrt{S_{odd}}}\exp\left(\pm \int^x_{x_0}S_{odd} dx\right).
\end{eqnarray}
The above WKB expansion is usually divergent but is Borel-summable.
Namely, one can consider 
\begin{eqnarray}
\psi_\pm &\rightarrow&\Psi_\pm\equiv\int^\infty_{\mp s(x)}e^{-y\eta}\psi_\pm^B(x,y)dy,\\
&&s(x)\equiv \int^x_{x_0}S_{-1}(x)dx,
\end{eqnarray}
where the $y$-integral is parallel to the real axis.
The Stokes phenomenon in the EWKB is explained using the Airy function
($Q(x)=x$) near the turning points.
If one defines the Stokes line starting from the turning point
at $x=0$ as 
\begin{eqnarray}
\mathrm{Im} [s(x)]=0,
\end{eqnarray}
The Stokes lines are the solutions of
\begin{eqnarray}
\mathrm{Im} [s(x)]&=& \mathrm{Im} \left[\int^x_0 x^{1/2}dx \right]\nonumber\\
&=&\mathrm{Im} \left[\frac{2}{3}x^{3/2}\right]=0,
\end{eqnarray}
which can be written as Fig.\ref{fig_airy}.
If $\mathrm{Re}[s(x)]>0$, $\psi_+$ is dominant on the Stokes line, while
if $\mathrm{Re}[s(x)]<0$, $\psi_-$ is dominant.

\begin{figure}[t]
\centering
\includegraphics[width=1.0\columnwidth]{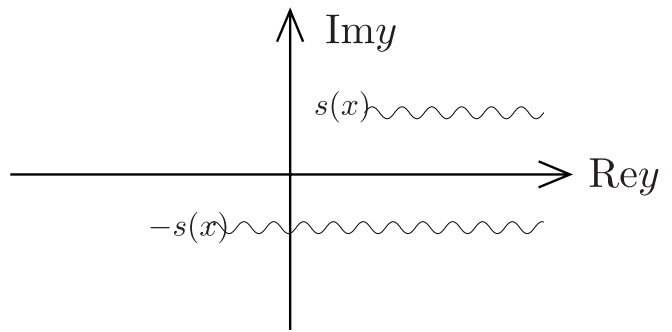}
 \caption{The wavy lines are the paths of integration in the Borel
 summation.
The paths overlap when $\mathrm{Im}[s]=0$.}
\label{fig_borelintegral}
\end{figure}

The paths of integration are given in Fig. \ref{fig_borelintegral}.
Note that the paths overlap on the Stokes line, since the Stokes lines
are defined as the solutions of $\mathrm{Im} [s(x)]=0$.
Therefore, the paths may develop additional contributions when $x$
goes across the Stokes line.
This is called the Stokes phenomenon.

Using the above idea, one can develop the following connection formulae:
\begin{itemize}
\item Crossing the $\psi_+$-Dominant Stokes line with anticlockwise rotation (seen from the turning
      point)
\begin{eqnarray}
\Psi_+&\rightarrow& \Psi_+ +i\Psi_-\\
\Psi_-&\rightarrow& \Psi_-
\end{eqnarray}
\item Crossing the $\psi_-$-Dominant Stokes line with anticlockwise rotation (seen from the turning
      point)
\begin{eqnarray}
\Psi_-&\rightarrow& \Psi_- +i\Psi_+\\
\Psi_+&\rightarrow& \Psi_+
\end{eqnarray}
\item Inverse rotation given a minus sign in front of $i$.
\end{itemize}

\begin{figure}[t]
\centering
\includegraphics[width=1.0\columnwidth]{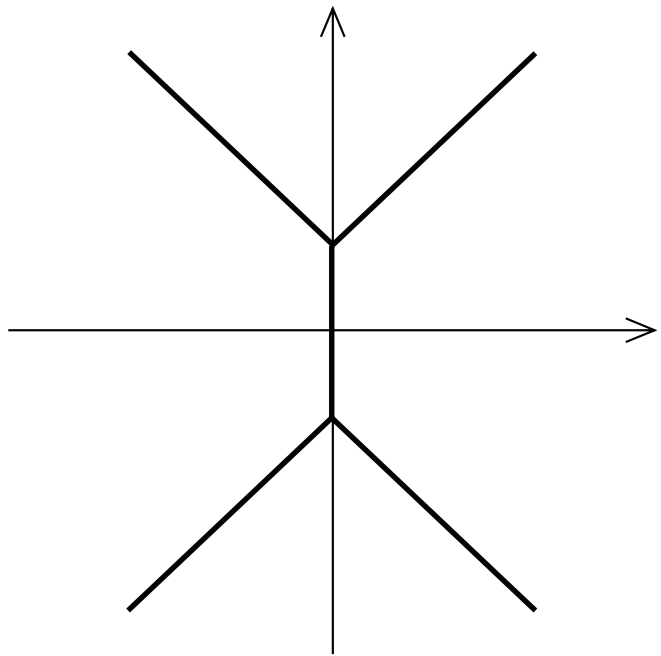}
 \caption{The Stokes lines for the inverted quadratic potential
 ($E>V$). The branch points are the complex ``turning points.''}
\label{fig_stokes_invquad}
\end{figure}

\begin{figure}[t]
\centering
\includegraphics[width=1.0\columnwidth]{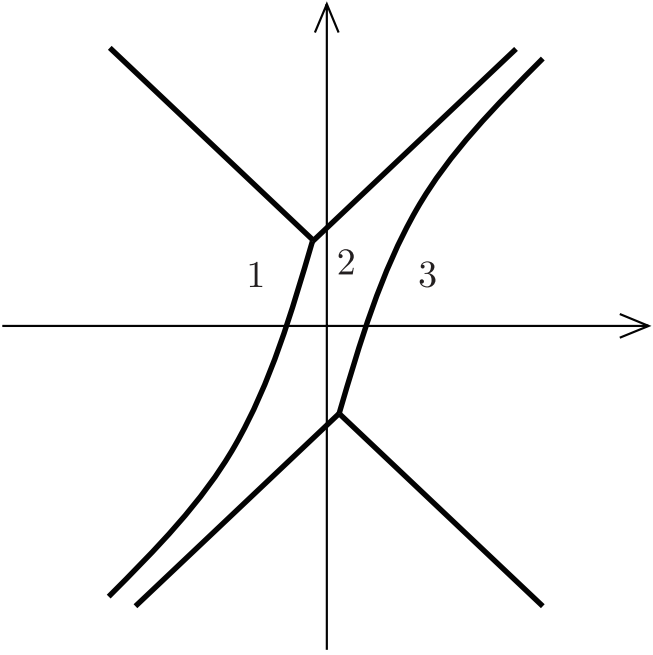}
 \caption{Split by $E\rightarrow E +i \epsilon$.}
\label{fig_stokes_invquad-deltae}
\end{figure}

Let us use these simple formulae to solve the scattering problem 
by the inverted quadratic potential ($E>0$).
The Stokes lines are given by Fig.\ref{fig_stokes_invquad}, which has
an MTP.
The degeneracy of the Stokes line can be solved by introducing imaginary
parameters such as $\eta\rightarrow \eta\pm i \eta_\epsilon$ or
$E\rightarrow E\pm i\epsilon$.
See Fig.\ref{fig_stokes_invquad-deltae}.
However, if one ignores the normalization factor, the $\pm$ splittings are not
consistent.
Fortunately, in the EWKB, the factor can be calculated
explicitly without relying on physical requirements\cite{Silverstone:2008}.
For the normalization factor, we have 
\begin{eqnarray}
\psi_{\pm}&=&\frac{1}{\sqrt{S_{odd}}}
\exp\left(\pm \int^x_{x_0}S_{odd} dx\right)\nonumber\\
&=&\psi^{(\infty)}_\pm 
\exp\left(\pm \int^x_{x_0}\left(S_{odd}-\eta S_{-1}\right) dx\right),\\
\psi^{(\infty)}_\pm&=&\frac{1}{\sqrt{S_{odd}}}
\exp\left(\pm \int^x_{x_0}S_{-1} dx\right) \nonumber\\
&&\times
\exp\left(\pm \int^x_{\infty}\left(S_{odd}-\eta S_{-1} \right)dx\right).
\end{eqnarray}
Among them, what is not trivial is 
\begin{eqnarray}
\exp\left(\pm \int^x_{x_0}\left(S_{odd}-\eta S_{-1}\right) dx\right).
\end{eqnarray}
This gives for $Q(x)=\lambda-\frac{x^2}{4}$,
\begin{eqnarray}
&&2\int^x_{2\sqrt{\lambda}}\left(S_{odd}-\eta S_{-1}\right) dx\nonumber\\
&=&\sum^\infty_{n=1} \frac{2^{1-2n}-1}{2n(2n-1)}B_{2n}(i\eta\lambda)^{1-2n},
\end{eqnarray}
where $B_{2n}$ is the Bernoulli number.
Because of this, one has the gap given by
\begin{eqnarray}
\Psi_+^{(\mathrm{Im}\lambda<0)}&=&\sqrt{1+e^{-2\pi\lambda \eta}}
\Psi_+^{(\mathrm{Im}\lambda>0)}.
\end{eqnarray}
The calculation can be generalized to give 
the factor appearing on both sides of more generic MTP\cite{Aoki:2009}.
In this paper, based on these mathematical results, we derive the
MTP factor from the simple consistency relation.

Analytic continuation of the integration requires identification of the
sheet on which the path is placed.
Since a cut always appears from a turning point, the sign of the integration
will be inverted when the path goes around a turning point.
Therefore, if the integration path going back and forth to the turning
point makes a turn around the turning point, the integration
does not cancel out.
This property is often used to define the integration in the EWKB.

\subsection{Exact WKB for $E>0$ inverted quadratic potential} \label{sec:EWKB_E_pos_quardra}
Now we can calculate the connection formula for the inverted quadratic
potential using the EWKB.
We start with
\begin{eqnarray}
\frac{d^2 \psi}{dt^2}+\left[k^2+g^2_2v^2t^2\right]\psi=0 \label{eq:eom_inverted_potential}
\end{eqnarray}
The initial time $t$ starts from the region 1 in Fig.\ref{fig_stokes_invquad-deltae}
and it connects to the right region 3 through the region 2.
First, we consider the WKB solution whose integral starts from a turning point $t=i
\frac{k}{g_2v}\equiv t_*^+$.
This point can be shared both in region 1 and 2.
We are crossing the $\Psi_-$ dominant Stokes line anticlockwise, and
the connection formula becomes
\begin{eqnarray}
\left(
\begin{array}{c}
\Psi_+^{(2u)}\\
\Psi_+^{(2u)}
\end{array}
\right)
&=&
\left(
\begin{array}{cc}
1&0\\
i&1
\end{array}
\right)
\left(
\begin{array}{c}
\Psi_+^{(1u)}\\
\Psi_+^{(1u)}
\end{array}
\right).
\end{eqnarray}
Here, $(1u)$ means ``in the region 1, integration starts from the
upper turning point at $t=t_*^+$''.
The starting point of the integration has to be changed when one moves
from the region 2 to 3.
We define
\begin{eqnarray}
K_{ud}&\equiv&  \int^{t_*^-}_{t_*^+}S_{odd}dt \label{eq:K_ud}
\end{eqnarray}
where $t_*^-\equiv -i\frac{k}{g_2v}$ and 
\begin{eqnarray}
\left(
\begin{array}{c}
\Psi_+^{(2u)}\\
\Psi_+^{(2u)}
\end{array}
\right)
&\rightarrow&
\left(
\begin{array}{cc}
e^{K_{ud}}&0\\
0&e^{-K_{ud}}
\end{array}
\right)
\left(
\begin{array}{c}
\Psi_+^{(2d)}\\
\Psi_+^{(2d)}
\end{array}
\right).
\end{eqnarray}
Now the path crosses the $\Psi_+$ dominant Stokes line clockwise, which
gives the connection formula
\begin{eqnarray}
\left(
\begin{array}{c}
\Psi_+^{(3d)}\\
\Psi_+^{(3d)}
\end{array}
\right)
&=&
\left(
\begin{array}{cc}
1&-i\\
0&1
\end{array}
\right)
\left(
\begin{array}{c}
\Psi_+^{(2d)}\\
\Psi_+^{(2d)}
\end{array}
\right).
\end{eqnarray}
Let us derive the normalization factor using a simple consistency
relation.
For the normalization factor $N$, we have\cite{Silverstone:2008} 
\begin{eqnarray}
\Psi_{\pm}^{1\infty}&=&N^{\mp}\Psi_{\pm}^{(1+)}\\
\Psi_{\pm}^{3\infty}&=&N^{\mp}\Psi_{\pm}^{(3-)},
\end{eqnarray}
which gives the connection formula from 1 to 3 as
\begin{eqnarray}
\left(
\begin{array}{c}
\Psi_+^{(3\infty)}\\
\Psi_+^{(3\infty)}
\end{array}
\right)
&=&
U_{N} U_{+c} U_{ud} U_{-a} U_{N}
\left(
\begin{array}{c}
\Psi_+^{(1\infty)}\\
\Psi_+^{(1\infty)}
\end{array}
\right),
\end{eqnarray}
where
\begin{eqnarray}
U_{+a}&\equiv&\left(
\begin{array}{cc}
1&i\\
0&1
\end{array}
\right)\\
U_{+c}&\equiv&\left(
\begin{array}{cc}
1&-i\\
0&1
\end{array}
\right)\\
U_{-a}&\equiv&\left(
\begin{array}{cc}
1&0\\
i&1
\end{array}
\right)\\
U_{-c}&\equiv&\left(
\begin{array}{cc}
1&0\\
-i&1
\end{array}
\right)\\
U_{ud}&\equiv&\left(
\begin{array}{cc}
e^{K_{ud}}&0\\
0&e^{-K_{ad}}
\end{array}
\right)\\
U_{N}&\equiv&\left(
\begin{array}{cc}
N&0\\
0&N^{-1}
\end{array}
\right).
\end{eqnarray}
The result is 
\begin{eqnarray}
\left(
\begin{array}{cc}
N^2 \left(e^{-K_{ud}} +e^{K_{ud}} \right) & -i e^{-K_{ud}} \\
 i e^{-K_{ud}} & \frac{e^{-K_{ud}}}{N^2} \\
\end{array}
\right).
\end{eqnarray}
To make the diagonal elements consistent, we have 
\begin{eqnarray}
|N|^{-2}=\sqrt{1+e^{2K_{ud}}}.
\end{eqnarray}
Finally, the connection matrix becomes
\begin{eqnarray}
\left(
\begin{array}{cc}
\sqrt{1+e^{-2K_{ud}}} & -i e^{-K_{ud}} \\
 i e^{-K_{ud}} & \sqrt{1+e^{-2K_{ud}}} \\
\end{array}
\right),
\end{eqnarray}
where the phase of $N$, which is calculable in the EWKB but not
calculable from the simple consistency relation only, has been neglected for
simplicity.

\begin{figure}[t]
\centering
\includegraphics[width=1.0\columnwidth]{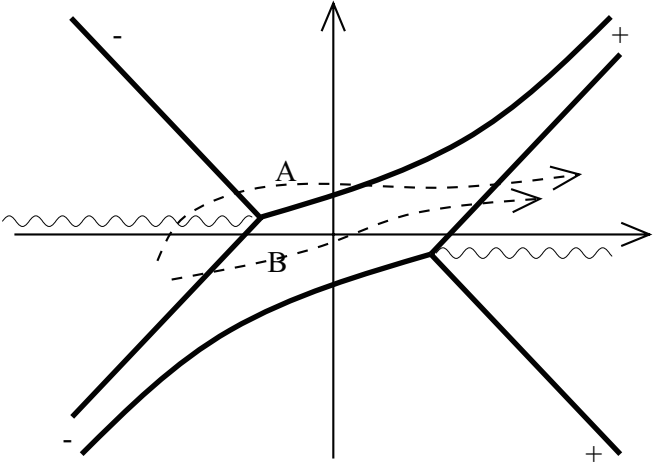}
 \caption{The Stokes lines for the inverted quadratic potential with
 $E<0$. The wavy lines are the cuts.}
\label{fig_stokesELdelE}
\end{figure}

The same calculation is possible for $E<0$, in which classical turning
points appear. 
For later convenience, we explicitly show the Stokes lines in
Fig.\ref{fig_stokesELdelE}.
The connection matrix for crossing the cut is
\begin{eqnarray}
U_{cc}&\equiv&\left(
\begin{array}{cc}
0&i\\
i&0
\end{array}
\right)\\
U_{ca}&\equiv&\left(
\begin{array}{cc}
0&-i\\
-i&0
\end{array}
\right),
\end{eqnarray}
where $cc$ and $ca$ denote the clockwise and the anticlockwise motion
across the cut.
For the path A, the connection matrix is
\begin{eqnarray}
&&U_N U_{+c} U_{LR} U_{+c} U_{-c} U_{cc} U_N\nonumber\\
&=&\left(
\begin{array}{cc}
 \left(e^{-K_{LR}} +e^{ K_{LR}}\right) N^2 & -i e^{-K_{LR}} \\
 i e^{-K_{LR}} & \frac{e^{-K_{LR}}}{N^2} \\
\end{array}
\right)
\end{eqnarray}
where
\begin{equation}
 U_{LR}\equiv\left(\begin{array}{cc}e^{K_{LR}} & \\ & e^{-K_{LR}} \end{array}\right)
\end{equation}
and
\begin{equation}
 K_{LR}\equiv\int_{t_*^L}^{t_*^R}S_{odd}dt, \quad t_*^L \equiv -\frac{k}{g_2v}, \quad t_*^R \equiv \frac{k}{g_2v}.
\end{equation}
Again, the consistency relation gives
\begin{eqnarray}
|N|^{-2}=\sqrt{1+e^{2K_{LR}}}.
\end{eqnarray}
The result is
\begin{eqnarray}
\left(
\begin{array}{cc}
\sqrt{1+e^{-2K_{LR}}} & -i e^{-K_{LR}} \\
 i e^{-K_{LR}} & \sqrt{1+e^{-2K_{LR}}} \\
\end{array}
\right).
\end{eqnarray}
For the path B, one has 
\begin{eqnarray}
&&U_N U_{+c} U_{LR} U_{-a} U_N\nonumber\\
&=&\left(
\begin{array}{cc}
 \left(e^{-K_{LR}} +e^{ K_{LR}}\right) N^2 & -i e^{-K_{LR}} \\
 i e^{-K_{LR}} & \frac{e^{-K_{LR}}}{N^2} \\
\end{array}
\right),
\end{eqnarray}
which is giving the identical result with the path A.

Now we can calculate the connection matrix for more exotic potential
using the simple EWKB.

\section{EWKB for higher dimensional interaction}
\subsection{Quantum scattering for inverted quartic potential}
We start with the inverted quartic potential 
$Q(x)=E+x^4$, which has four (complex) turning points
\begin{eqnarray}
x_n=E^{\frac{1}{4}}e^{i\frac{(2n-1) \pi}{4}},\,\,\,\, (n=1,..4).
\end{eqnarray}
\begin{figure}[t]
\centering
\includegraphics[width=1.0\columnwidth]{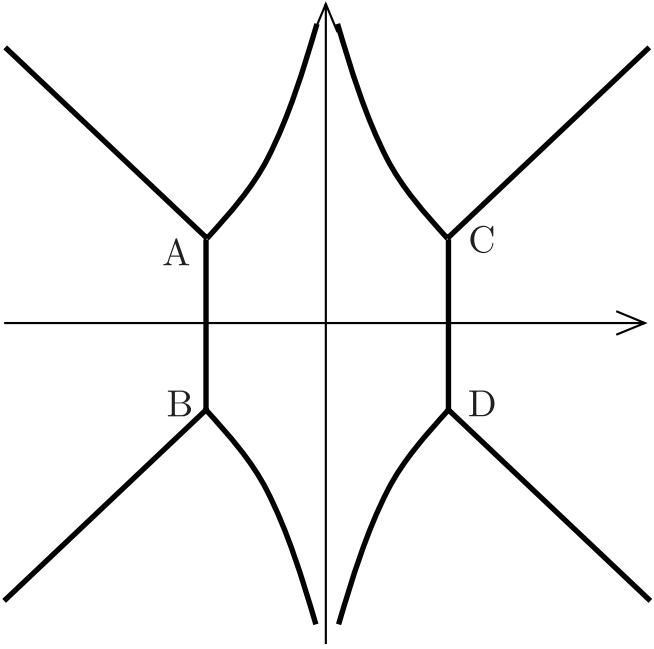}
 \caption{The Stokes lines for the inverted quartic potential with
 $E>0$.}
\label{fig_stokes_quartic}
\end{figure}
The Stokes lines are shown in Fig.\ref{fig_stokes_quartic}.
The previous calculation of the inverted quadratic potential with $E>0$
is suggesting that the connection matrix in this case is given by the product
of two matrices, each of which (including the MTP factors)
has to be calculated for each MTP.
Neglecting the phase parameters, the significant difference from the
conventional quadratic potential may only appear in the integration factors,
which are connecting the turning points $A-B$ and $C-D$. 

To make a more specific calculation, we consider the Lagrangian given by
\begin{eqnarray}
{\cal L}_\chi&=&\frac{1}{2}\partial_\mu \chi\partial^\nu
 \chi-\frac{1}{2}m_0^2\chi^2\nonumber\\
&&  -\frac{1}{2}g_2^2 \phi(t)^2\chi^2
-g_4^2 \frac{\phi(t)^4}{M_*^2}\chi^2,
\end{eqnarray}
where $M_*$ denotes the cut-off scale of the effective action.
One may take $M_*$ as large as the GUT(Grand Unified Theory) scale or
the Planck scale.
The effective mass of $\chi$ is given by
\begin{eqnarray}
m_\chi^2(t)&=&m_0^2 +g^2_2 \phi(t)^2+g_4^2\frac{\phi(t)^4}{M_*^2}.
\end{eqnarray}
The equation of motion is
\begin{eqnarray}
\frac{d^2 \chi}{dt^2}+\left[k^2+m^2_\chi(t)\right]\chi=0.
\end{eqnarray}
Again, we assume that the approximation $\phi(t)\simeq vt$ is possible
when the particles are produced.
Then, we have the Schr\"odinger equation for the scattering problem with
the inverted quartic potential, which is given by
\begin{eqnarray}
\frac{d^2 \chi}{dt^2}+\left[k^2+m_0^2+g^2_2v^2 t^2 
+\frac{g_4^2 v^4}{M_*^2}t^4\right]\chi&=&0. \label{eq:eom_chi_k}
\end{eqnarray}
Here the ``energy'' and the ``potential'' are given by
\begin{eqnarray}
E_k&=&k^2+m_0^2\\
V(t)&=&-\left(g_2^2v^2\right)t^2-\frac{g_4^2 v^4}{M_*^2}t^4.
\end{eqnarray}
According to the calculation we have given for the quadratic potential,
we translate
\begin{eqnarray}
z&=&\left(\frac{g_4^2v^4}{M_*^2}\right)^{\frac{1}{6}}t\\
&&a_4^2\equiv\left(\frac{g_4^2v^4}{M_*^2}\right)^{\frac{1}{3}}
=v\left(\frac{g_4^2v}{M_*^2}\right)^{\frac{1}{3}}
\end{eqnarray}
to make these quantities dimensionless, we have 
\begin{eqnarray}
V(z)&=&-z^4-g_2^2\left(\frac{M_*^2}{g_4^2v}\right)^{\frac{2}{3}}\\
E_k&=&\frac{k^2+m_0^2}{a_4^2}\equiv \kappa_4^2. \label{eq:def_kappa_4}
\end{eqnarray}
Defining $\delta_4\equiv
\left(\frac{g_4^2v}{M_*^2}\right)^{\frac{1}{3}}\ll 1$,
we have
\begin{eqnarray}
V(z)&=&-z^4- \frac{g_2^2}{\delta_4^{2}}z^2 \\
E_k&=&\kappa_4^2=\frac{k^2+m_0^2}{v\delta_4}.
\end{eqnarray}

\begin{figure}[t]
\centering
\includegraphics[width=1.0\columnwidth]{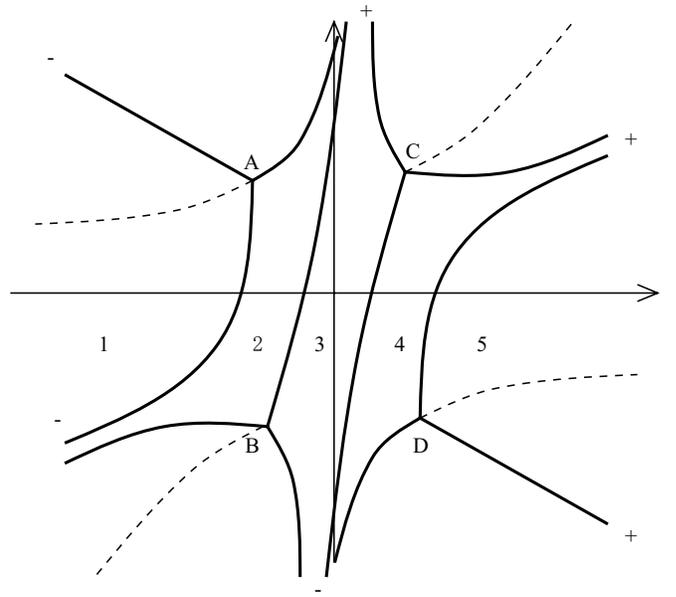}
 \caption{Split Stokes lines of MTPs in the simple quartic potential.}
\label{fig_stokes-simplequaltic}
\end{figure}

\begin{figure}[t]
\centering
\includegraphics[width=1.0\columnwidth]{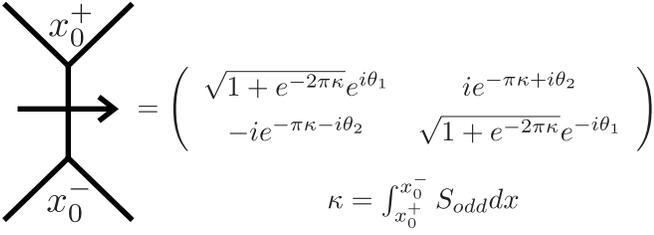}
 \caption{An MTP structure and its integration factor.}
\label{fig_stokes-parts}
\end{figure}

We start with the calculation for $g_2=0$ (pure quartic).
One can find a more explicit calculation in Ref.\cite{Voros:1983}, in
which the Stokes lines are shown to split into the form given in
Fig.\ref{fig_stokes-simplequaltic}.
Here, what we want is not the exact formula but an estimation of the
elements of the connection matrix.
In this case, the parameter which determines the magnitude of particle
creation is the integral connecting MTP, which is shown in Fig.\ref{fig_stokes-parts}.
For $Q(x)=V(x)-E_k=-x^4-\kappa_4^2$, the integration factor $K_{AB}$
connecting
the turning points at $(t_A=\sqrt{\kappa_4}e^{i\frac{3\pi}{4}})$
and $(t_B=\sqrt{\kappa_4}e^{i\frac{5\pi}{4}})$
becomes
\begin{eqnarray}
K_{AB}&=&2\int^B_A\sqrt{-x^4-\kappa_4^2}dx\\
&\simeq&  1.2 \kappa_4^{\frac{3}{2}}. \label{eq:def_K_AB}
\end{eqnarray}
If we define the critical parameter $k_*$ by
\begin{eqnarray}
  1.2 \kappa_4^{\frac{3}{2}}\simeq 1,
\end{eqnarray}
we have 
\begin{eqnarray}
k_*&\simeq&\sqrt{v}\times \left(\frac{\delta_4}{1.3}\right)^{1/2}
\end{eqnarray}
in a case of $k_*\gtrsim m_0$.
Typically, we have
\begin{eqnarray}
\delta_4\simeq 10^{-5/3}\left(\frac{g_4}{0.1}\right)^{1/3}
\left(\frac{\sqrt{v}}{10^{16} GeV}\right)^{2/3}
\left(\frac{10^{18} GeV}{M_*}\right)^{2/3},\nonumber\\
\end{eqnarray}
which gives
\begin{eqnarray}
k_*&\sim&0.1\sqrt{v}.
\end{eqnarray}
This result shows that despite the significant suppression by the
cut-off scale appearing in
the higher interaction, the integration factor may not introduce significant suppression to the
particle production.
From this result, one can see that the typical (explicit) interaction
is not always necessary for the cosmological preheating scenario, since
higher (Planck-suppressed) terms can cause significant particle
production after inflation.
This result is consistent with our earlier works\cite{Enomoto:2013mla, Enomoto:2014hza}.
Note also that the result is obtained for a single process, which means
that parametric resonance is not assumed to obtain significant particle
production by the Planck-suppressed interaction.
This result may affect the conventional estimation of cosmological moduli production,
since any (hidden) particle can have Planck-suppressed interaction.

\subsection{Classical scattering for inverted double-well potential}
Previously we analyzed the cosmological effect of classical turning points
of the inverted quadratic potential and mentioned the
``amplification'', which had specific meanings.
Here we consider amplification by the classical turning points of the higher interaction.

\begin{figure}[t]
\centering
\includegraphics[width=1.0\columnwidth]{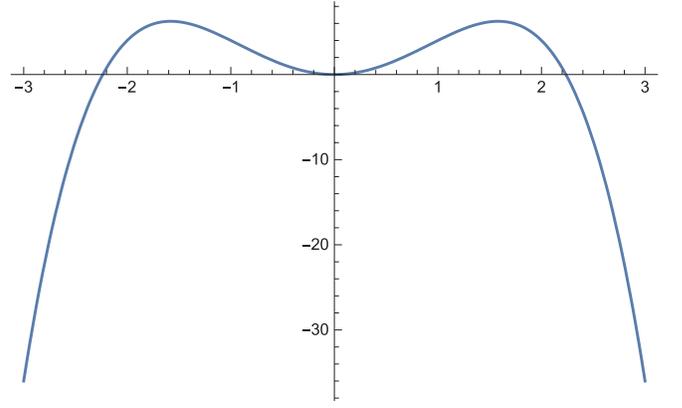}
 \caption{Inverted double-well potential of $V=5x^2-x^4$.}
\label{fig_invdouble}
\end{figure}

To introduce scattering by a classical turning point, we consider a negative coefficient for
the quadratic term (i.e., $g_2^2$ is replaced by $-g^2_2$).
Just for simplicity, we put $m_0=0$.
Typically, the potential becomes an inverted double-well form, which is
shown in Fig.\ref{fig_invdouble}.
We consider the scattering problem of the Schr\"odinger equation with
\begin{eqnarray}
V(z)&=&-z^4+ \frac{g_2^2}{\delta_4^{2}}z^2 \\
E_k&=&\kappa_4^2=\frac{k^2}{v\delta_4}.
\end{eqnarray}
The two bumps of the potential are placed at
\begin{eqnarray}
z=\pm \frac{g_2}{\sqrt{2}\delta_4}
\end{eqnarray}
and the height is given by
\begin{eqnarray}
V\left(\pm \frac{g_2}{\sqrt{2}\delta_4}\right)=\frac{1}{4}\frac{g_2^4}{\delta_4^4}.
\end{eqnarray}
Therefore, the ``tachyonic'' region appears when $E_k<V(z)$.
If we define $k_0$ by 
\begin{eqnarray}
\frac{k_0^2}{v\delta_4} &=&\frac{1}{4}\frac{g_2^4}{\delta_4^4},
\end{eqnarray}
we find
\begin{eqnarray}
k_0&=&\frac{1}{2}v\frac{g_2^2}{\delta_4^{3/2}}\nonumber\\
&\simeq&\sqrt{v} \times 10^{3/2} \times 
\left(\frac{g_2}{10^{-2}}\right)^2
\left(\frac{10^{-11/3}}{\delta_4}\right)^{\frac{3}{2}}.
\end{eqnarray}
From the above result, one can confirm in this case that there can be
the ``amplification'' of the particle production by the classical
turning points.
This is our new result in this paper.

\section{Fermions}
Typically, equation of motion of a fermion is described by a pair of first-order differential
equations.
We are going to see how higher terms appear in a fermionic model and
discuss its cosmological implications.
Note that for fermions, higher (quartic) potential may appear in the
corresponding Schr\"odinger equation without introducing higher
dimensional interaction to the theory. 
We are going to discuss the above last topic first.

\subsection{Landau-Zener model and kinematic particle creation in cosmology}
\begin{figure}[ht]
\centering
\includegraphics[width=0.9\columnwidth]{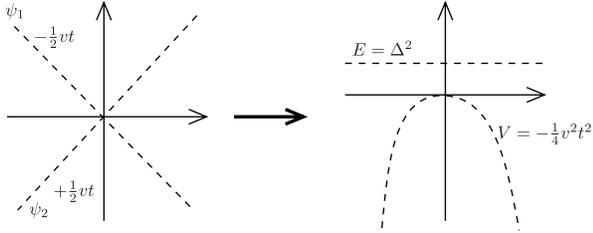}
 \caption{Left : State crossing in the Landau-Zener model.
Right : After decoupling the equations, the state crossing of the
 Landau-Zener model corresponds to the scattering problem of the
 Schr\"odinger equation. }
\label{fig_LZtoPH}
\end{figure}

We first review the Landau-Zener model and explain its relation between
cosmological particle creation.
We start with a pair of first-order differential equations below
\begin{eqnarray}
i\hbar\frac{d}{dt}\left(
\begin{array}{c}
\psi_1\\
\psi_2
\end{array}
\right)&=&\left(
\begin{array}{cc}
-\frac{v}{2}t& \Delta \\
 \Delta& +\frac{v}{2}t 
\end{array}
\right)
\left(
\begin{array}{c}
\psi_1\\
\psi_2
\end{array}
\right),
\end{eqnarray}
where $v>0$ and $\Delta$ is real.
Decoupling the equations, we have
\begin{eqnarray}
\left[\frac{d^2}{dt^2}+\frac{1}{\hbar^2}\left(\Delta^2+\frac{1}{4}v^2t^2\right)-\frac{i}{\hbar}\frac{v}{2}\right]\psi_1&=&0 \\
\left[\frac{d^2}{dt^2}+\frac{1}{\hbar^2}\left(\Delta^2+\frac{1}{4}v^2t^2\right)+\frac{i}{\hbar}\frac{v}{2}\right]\psi_2&=&0.
\end{eqnarray}
These are the Schr\"odinger equations with
$V=-\frac{1}{\hbar^2}\cdot\frac{1}{4}v^2 t^2$ and
$E=\frac{1}{\hbar^2}\cdot\Delta^2\pm \frac{i}{\hbar}\frac{v}{2}$, where
the energy is shifted by the imaginary part, which has an
extra $\hbar$.
Therefore, we have to consider a generalized $Q(z,\eta)$ for the EWKB
analysis.
The EWKB analysis of $Q(z,\eta)$ has been discussed in
Ref.\cite{Aoki:1993}, in which it has been suggested that 
both the turning points and the
Stokes lines have to be calculated for the leading terms
if one expands $Q(z,\eta)$ with $\eta^{-n}$.
Hence, we analyse the Stokes lines by
\begin{eqnarray}
\left[\frac{d^2}{dt^2}+\Delta^2+\frac{1}{4}v^2t^2\right]\psi&=&0,
\end{eqnarray}
where the imaginary part has been removed and $\hbar=1$ is chosen.
See also Fig.\ref{fig_LZtoPH}.

Let us first calculate the connection matrix using the exact solution
based on the Weber function.
Defining $z=i\sqrt{v} e^{i\pi/4}t$ ($z^2=-ivt^2$), we have
\begin{eqnarray}
\left[\frac{d^2}{dz^2}+\left(n+\frac{1}{2}-\frac{1}{4}z^2\right)\right]\psi_1(z)&=&0.
\end{eqnarray}
Here we defined
\begin{eqnarray}
n&\equiv&i\frac{\Delta^2}{v}.
\end{eqnarray}
This is nothing but the standard equation of the Weber function.
The solutions are given by a pair of independent functions among 
$D_n(z), D_n(-z), D_{-n-1}(iz), D_{-n-1}(-iz)$.
According to the previous calculations, we can easily find the
connection matrix connecting the initial ($\psi^i$) to the end
($\psi^e$) states.
\begin{eqnarray}
\left(
\begin{array}{c}
\psi_1^e\\
\psi_2^e
\end{array}
\right)&=&\left(
\begin{array}{cc}
e^{-\pi \kappa}e^{i\theta_1}& -\sqrt{1-e^{-2\pi\kappa}}e^{i\theta_2} \\
\sqrt{1-e^{-2\pi\kappa}}e^{-i\theta_2} & e^{-\pi \kappa}e^{-i\theta_1}
\end{array}
\right)
\left(
\begin{array}{c}
\psi_1^i\\
\psi_2^i
\end{array}
\right),\nonumber\\
\end{eqnarray}
where
\begin{eqnarray}
\kappa&\equiv&\frac{\Delta^2}{v}.
\end{eqnarray} 
Note however that $\psi_{1,2}$ are not the adiabatic states.
Therefore, one can see that the ``translation'' becomes significant
when $v\rightarrow 0$.
To avoid confusion, the above matrix has to be given for the adiabatic
states, which are not identical to
$\psi_{1,2}$.\footnote{Note that usually the ``adiabatic
energy'' is given by $E_\pm=\pm\sqrt{\Delta^2+v^2t^2/4}$, which do not
intersect.}
We thus define the adiabatic states to have the matrix given by
\begin{eqnarray}
\left(
\begin{array}{c}
\Psi_1^+\\
\Psi_2^+
\end{array}
\right)&=&\left(
\begin{array}{cc}
 \sqrt{1-e^{-2\pi\kappa}} &e^{-\pi \kappa}\\
e^{-\pi \kappa} &-\sqrt{1-e^{-2\pi\kappa}}
\end{array}
\right)
\left(
\begin{array}{c}
\Psi_1^-\\
\Psi_2^-
\end{array}
\right),\nonumber\\
\end{eqnarray}
where phase parameters are neglected.

Seeing the relationship between the Landau-Zener model and the scattering
for the Schr\"odinger equation, one will find that the original (linear)
$D_1\equiv - vt, D_2\equiv +vt$ in the diagonal elements are giving
the quadratic potential $-\frac{1}{4}v^2t^2$ for the Schr\"odinger
equation.
Finally, the problem is equivalent to the scattering with
$E=\Delta^2$ and $V(t)\equiv-\frac{1}{4}v^2t^2$.
(See also Fig.\ref{fig_LZtoPH}.)
The discussion of the EWKB suggests that the imaginary part of $E$,
which is the next order to $\Delta^2$,
does not change the connection matrix because it does not change the
Stokes lines.

\begin{figure}[ht]
\centering
\includegraphics[width=0.9\columnwidth]{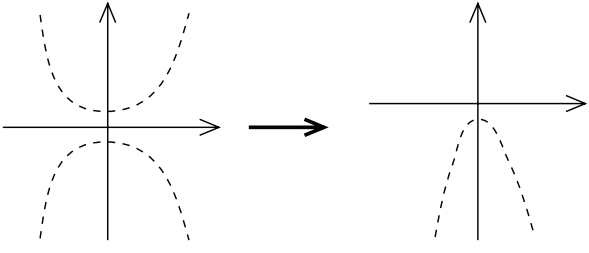}
 \caption{Left : States in the Landau-Zener model, when two states do
 not cross. Right : The situation is
 shown for the decoupled (Schr\"odinger) equation.
 The left is simply showing the diagonal elements, not the
 adiabatic energy.}
\label{fig_LZhigher}
\end{figure}

Since cosmological particle production deals with oscillatory
phenomena, which means that the diagonal elemants of the
Landau-Zener model is an approximation of a sinusoidal function, the
intersection of the two states may not be approximated by $\pm vt$.
To explain the situation, we prepared Fig.\ref{fig_LZhigher}.
As is shown in Fig.\ref{fig_LZhigher}, two states may not cross 
but just approach for a short time.
In this case, the diagonal elements are not approximated by 
$D_1=-D_2=-vt/2$ but by $D_1=-D_2=-\epsilon -at^2$.
Then, one has to calculate the transition when the velocity vanishes at
$t=0$.
Since the velocity vanishes, one cannot use the original argument.
We are going to solve this problem by using the EWKB.
As is shown in Fig.\ref{fig_LZhigher}, states (the original states $\psi_{1,2}$) are not crossing but
approach to the distance $2\epsilon$, and then moves away 
from each other.

In this case, in contrast to the conventional Landau-Zener model, one can
identify the original states as the adiabatic states, since there is no
crossing of the states.
(i.e, we do not have to consider $\psi_{1,2}\rightarrow\Psi_{1,2}$.)

\begin{figure}[ht]
\centering
\includegraphics[width=0.9\columnwidth]{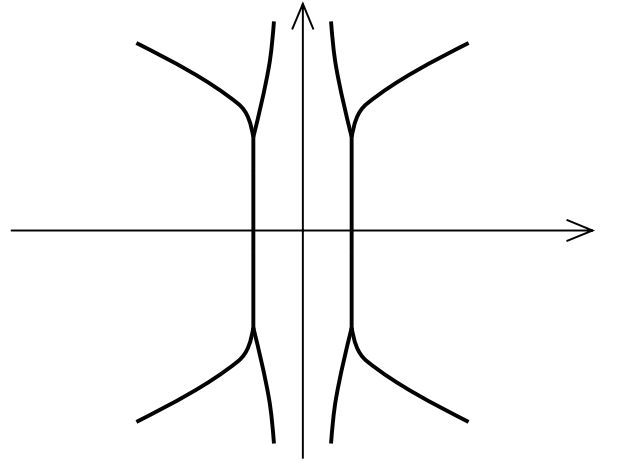}
 \caption{The Stokes lines for the fermion preheating when ``crossing''
 is not significant.}
\label{fig_stokes_LZhigher}
\end{figure}

We start with the equation given by
\begin{eqnarray}
i\frac{d}{dt}\left(
\begin{array}{c}
\psi_1\\
\psi_2
\end{array}
\right)&=&\left(
\begin{array}{cc}
-(a t^2+\epsilon)& \Delta \\
 \Delta& at^2+\epsilon 
\end{array}
\right)
\left(
\begin{array}{c}
\psi_1\\
\psi_2
\end{array}
\right).
\end{eqnarray}
Combining the equations, we find
\begin{eqnarray}
\left[\frac{d^2}{dt^2}+\left(\Delta^2-i(2at)\right)+\left(at^2+\epsilon\right)^2\right]\psi_1&=&0.
\end{eqnarray}
Again, the imaginary part should have an additional $\hbar$ and can be
neglected, since our focus is the connection matrix calculated from the
Stokes lines.
We thus have
\begin{eqnarray}
\label{eq-fermion1}
\left[\frac{d^2}{dt^2}+\Delta^2+\left(at^2+\epsilon\right)^2\right]\psi&=&0.
\end{eqnarray}
for the calculation.
This equation is nothing but the Schor\"odinger equation for the
scattering problem with the inverted quartic potential
\begin{eqnarray}
V(t)&=&-(at^2+\epsilon)^2-\Delta^2<0
\end{eqnarray}
and the energy $E=0$.
Now the connection matrix can be calculated along the real axis, as we
have shown for the bosonic preheating scenario with the higher intersection.
For the EWKB, we have 
%\begin{eqnarray}
%Q(z)&=&-\Delta^2-\epsilon^2- a^2 z^4-2a\epsilon z^2.
%\end{eqnarray}
\begin{eqnarray}
Q(t)&=&-\Delta^2-\epsilon^2- a^2 t^4-2a\epsilon t^2.
\end{eqnarray}
The stokes lines for Eq.(\ref{eq-fermion1}) is given by
Fig.\ref{fig_stokes_LZhigher}, which shows that the connection matrix is
calculated from the two MTPs.
Again, the important factor that determines the particle production is 
the integral with respect to the MTPs.

We are going to examine these results from the viewpoint of 
cosmological particle production and discuss the physics 
behind them.

\subsection{Cosmological particle production of fermions.}
Since the fermionic preheating with a linear $\phi(t)$ is an old idea,
we carefully follow Ref.\cite{Peloso:2000hy} to avoid confusion.
We consider a Dirac fermion whose mass is given by
\begin{eqnarray}
m_D(t)&=& m_0 + g\phi(t),
\end{eqnarray}
where $m_D$ is assumed to be real.
The Dirac equation is given by
\begin{eqnarray}
(i\slashed{\partial}-m_D)\psi&=&0,
\end{eqnarray}
whose solution can be decomposed as
\begin{eqnarray}
\psi&=&\int\frac{d^3k}{(2\pi)^3}
e^{-i\boldsymbol k\cdot \boldsymbol x}\sum_s\left[
u^s_{\boldsymbol k}(t)a^s_{\boldsymbol k}
+v^s_{\boldsymbol k}(t)b^{s\dagger}_{-\boldsymbol k}\right].
\end{eqnarray}
Choosing the momentum along the third direction $k=k_z$,
and introducing $u_\pm$ following Ref.\cite{Peloso:2000hy},
one can obtain a two-component differential equation 
given by\footnote{According to \cite{Peloso:2000hy},
the representation of the gamma matrices are chozen as
\begin{equation}
 \gamma^0=\left(\begin{array}{cc}\mathbf{1} & \\ & -\mathbf{1} \end{array} \right), \quad
 \gamma^1=\left(\begin{array}{cc} & -i\sigma^2 \\ -i\sigma^2 & \end{array} \right),
\end{equation}
\begin{equation}
 \gamma^2=\left(\begin{array}{cc} & i\sigma^1 \\ i\sigma^1 & \end{array} \right), \quad
 \gamma^3=\left(\begin{array}{cc} & \mathbf{1} \\ -\mathbf{1} & \end{array} \right).
\end{equation}}
\begin{eqnarray}
\dot{u}_\pm&=&ik u_{\mp}\mp i m_D u_\pm.
\end{eqnarray}
The equation can be written in the following form
\begin{eqnarray}
i\frac{d}{dt}\left(
\begin{array}{c}
u_+\\
u_-
\end{array}
\right)&=&\left(
\begin{array}{cc}
m_0+g\phi(t)& -k \\
 -k & -m_0-g\phi(t)
\end{array}
\right)
\left(
\begin{array}{c}
u_+\\
u_-
\end{array}
\right),\nonumber\\
\end{eqnarray}
which is giving the starting point of the Landau-Zener model.
Combining the equations, we have\cite{Peloso:2000hy}
\begin{eqnarray}
\label{eq-dec}
\ddot{u}_\pm+\left[\omega^2\pm i\dot{m}_D\right]u_\pm=0,
\end{eqnarray}
where we defined $\omega(t)^2=k^2+m_D^2$.
In writing the Stokes lines, the imaginary
term $i\dot{m}_D$ has to be neglected, since it has 
an additional $\hbar$.
The typical preheating scenario considers significant particle creation near
the Enhanced Symmetric Point(ESP), where one expects $m_D=0$ and $v\ne 0$.
In this case, the motion during particle production is given by
$\phi(t)\simeq vt$, which leads to the state crossing.
Of course, this expectation is true when $m_0$ is small.
Even for large $m_0$, fermionic preheating expects $m_D=0$ when
$\phi(t_0)=-\frac{m_0}{g}$, but on such point the linear approximation
may not be a good approximation if $\phi(t_0)$ appears near the edge of
the oscillation, or even does not appear, as is already shown in Fig.\ref{fig_LZhigher}

Now we can examine ``the Landau-Zener model without the state crossing''
from the cosmological viewpoint.
If $m_0$ is large and the mass is given by $m_D(t)= M(1+\Gamma(t)\cos
m_\phi t)$, where $\Gamma(t)$ is introduced to describe the damping of
the oscillation, the particle creation may be
significant when the above (linear) approximation is no longer valid.
This may happen during the first oscillation.
In that case, one has to calculate the particle production with
\begin{eqnarray}
m_D&\simeq& \frac{M \Gamma m_\phi^2}{2} (t-t_0)^2 +M(1-\Gamma).
\end{eqnarray}
Assuming $\Gamma\simeq 1$ for simplicity, 
the corresponding Schr\"odinger equation has 
\begin{eqnarray}
E&=&k^2\nonumber\\
V(t)&=& -\frac{M^2 m_0^4}{4}(t-t_0)^4 \mp i M m_0^2(t-t_0).
\end{eqnarray}
Again, the particle production can be calculated in terms of the
scattering problem of the Schr\"odinger equation, but in this case the
quadratic potential is replaced by the quartic potential.
This means that the production of heavy fermions (such as right-handed
Majorana fermions or GUT particles decaying to generate baryon/lepton
number) could be affected by such higher-order contribution.\footnote{We
are not claiming that the contribution drastically changes the previous
results\cite{Greene:1998nh,Giudice:1999fb}, since the situation
considered in this paper is different from the previous calculation.}

In contrast to bosonic preheating, the appearance of classical turning
points is unlikely.
To make the classical scattering possible for the fermions, one has to
put the four turning points together on the real axis of $t$, 
and such a scenario seems unreasonable.

In the above, we considered a scenario in which two states do not
cross (or cross but the linear approximation is no longer valid), 
and found that the scenario leads to the scattering with higher
potential.
Alternatively, genuine higher dimensional interaction ($m_D\propto
\phi^n, n\ge 2$) can be used to
introduce higher potential.
The latter is what we have discussed previously for bosons.

The EWKB has been a powerful tool in understanding the physics behind
the equations, especially when it is difficult to find the exact
solution in terms of special functions.

\section{Conclusions and discussions}
In this paper, we examined cosmological particle production caused by
a time-dependent background using the EWKB and the Stokes phenomena.
Our focus was higher-dimensional interactions and the ``classical'' scattering
process.
The latter appears when the corresponding Schr\"odinger equation
develops classical turning points on the real axis of time ($t$).
First, we examined a typical preheating scenario and found that
amplification by the classical scattering is not possible. 
Then, using the EWKB, we developed simple calculational methods for
preheating with the higher-dimensional interaction, and found
that such (classical) amplification is indeed possible for
the higher-dimensional interaction.
Finally, we analyzed fermionic preheating mentioning the two distinctive
sources of the higher-dimensional interaction and the possibility of the classical turning
points.

In our future works, we are going to analyze in detail the conditions for
asymmetric particle production\cite{matsuda_to_appear, Dolgov:1996qq,Funakubo:2000us,Rangarajan:2001yu,
Enomoto:2017rvc,Enomoto:2018yeu}, using the ideas developed in this
paper.
Since baryogenesis requires B-violating interaction, and the interaction
has to play a significant role in generating the asymmetry, the equations
have to be multi-component (i.e, higher-order after decoupling).
From the mathematical side, the Stokes phenomena of such higher-order
equations have been analyzed for decades using the EWKB, in which virtual
turning points are found to be crucial\cite{Virtual:2015HKT}.
While on the cosmological side, at least when compared with the mathematical
developments of the same period, the discussions of particle production and
baryogenesis have been quite ad-hoc.
In our future works, we are going to solve the Stokes phenomena of
multi-component equations when interactions (i.e, off-diagonal elements)
are time-dependent, to show explicitly how the asymmetry appears in
cosmological particle creation with time-dependent
backgrounds\cite{matsuda_to_appear}.
The EWKB considered in this paper is useful for calculating the connection formulae of
complicated differential equations, which cannot be solved simply using special
functions, and such equations may often appear in the cosmological
calculation of the baryon asymmetry.

\section{Acknowledgment}
The authors would like to thank Nobuhiro Maekawa for collaboration on the early stages of this work.
SE  was supported by the Sun Yat-sen University Science Foundation.

\appendix

\section{Comparison with the steepest descent method} \label{sec:steepest_descent_method}
In this section, we are going to review the calculation of the steepest descent method
and compare it with the EWKB.
The steepest descent method was applied to the cosmological particle
production in Ref.~\cite{Chung:1998bt, Enomoto:2013mla, Enomoto:2014hza}.
One can easily check that these two are consistent with each other.

Let us consider solving the following equation of motion:
\begin{equation}
 \ddot{\chi}_k+\left[k^2+m_0^2+\frac{g_{2n}^2}{M_*^{2(n-1)}}(vt)^{2n}\right]\chi_k = 0. \label{eq:eom_chi_k_2}
\end{equation}
This equation reproduces Eq.(\ref{eq:eom_chi_k});
$n=1$ gives $g_4=0$  and $n=2$ gives $g_2=0$.
If the adiabatic condition, which corresponds to
\begin{equation}
 \frac{\dot{\omega}_k}{\omega_k^2} \ll 1
\end{equation}
where
\begin{equation}
 \omega_k\equiv \sqrt{k^2+m_0^2+\frac{g_{2n}^2}{M_*^{2(n-1)}}v^{2n}t^{2n}}
\end{equation}
is satisfied, eq.(\ref{eq:eom_chi_k_2}) has the (conventional) WKB
solution given by
\begin{equation}
 \chi_k = \frac{\alpha_k}{\sqrt{2\omega_k}} e^{-i\int_{-\infty}^t dt'\omega_k(t')}
   + \frac{\beta_k}{\sqrt{2\omega_k}} e^{+i\int_{-\infty}^t
   dt'\omega_k(t')} \label{eq:WKB-type_solution}
\end{equation}
where $\alpha_k, \beta_k$ are the complex constants that satisfies 
\begin{equation}
 |\alpha_k|^2 - |\beta_k|^2 = 1.
\end{equation}

The WKB solution may have time-dependent coefficients such as 
\begin{equation}
 \alpha_k\rightarrow\alpha_k(t), \qquad \beta_k\rightarrow\beta_k(t),
\end{equation}
where the additional constraint
\begin{equation}
 0 = \frac{\dot{\alpha}_k}{\sqrt{2\omega_k}} e^{-i\int_{-\infty}^t dt'\omega_k(t')}
   + \frac{\dot{\beta}_k}{\sqrt{2\omega_k}} e^{+i\int_{-\infty}^t
   dt'\omega_k(t')}
\end{equation}
has to be satisfied.
Substituting the representation (\ref{eq:WKB-type_solution}) into (\ref{eq:eom_chi_k_2}) with the above constraint,
one can obtain
\begin{eqnarray}
 \dot{\alpha}_k &=& \beta_k \frac{\dot{\omega}_k}{2\omega_k} e^{+2i\int_{-\infty}^t dt' \omega_k},\\
 \dot{\beta}_k &=& \alpha_k \frac{\dot{\omega}_k}{2\omega_k}
  e^{-2i\int_{-\infty}^t dt' \omega_k}. \label{eq:eom_beta}
%\label{eq:EOM_of_beta}
\end{eqnarray}
Since we are considering a zero-particle state as the initial state,
the initial conditions are given by
\begin{equation}
 \alpha_k(-\infty)=1, \qquad \beta_k(-\infty)=0.
\end{equation}
At $t=+\infty$, $\beta_k$ may be non-zero if particle production happens around $t=0$.
Assuming that $|\beta_k(+\infty)|^2\ll1$, one can approximate
$\alpha_k\sim 1$ during the whole process.
Then, $\beta$ at the final state can be estimated from (\ref{eq:eom_beta}) as
\begin{equation}
 \beta_k(+\infty) = \int_{-\infty}^{+\infty} dt \: \frac{\dot{\omega}_k(t)}{2\omega_k(t)}
  \exp{\left[ -2i \int_{-\infty}^t dt' \omega_k(t') \right]}. \label{eq:solution_of_beta}
\end{equation}

\begin{figure}[t]
 \begin{center}
  \includegraphics[scale=0.8]{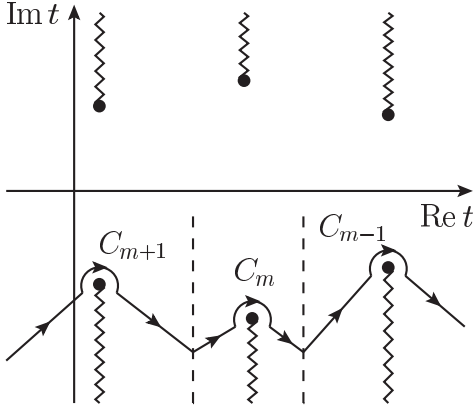}
  \caption{The path of integration on the $t$ complex plane.
   (This figure is the same as the one used in Ref.\cite{Enomoto:2013mla}.)
   The dots show the poles which are also saddle points,
  and the wavy lines show the branch cuts.}
  \label{fig:path}
 \end{center}
\end{figure}

Now, one can evaluate the integral of Eq.(\ref{eq:solution_of_beta}) using the steepest descent method.
To understand the integral, we focus on 
\begin{equation}
 \frac{\dot{\omega}_k}{2\omega_k} = \frac{n}{2}\frac{t^{2n-1}}{t_k^{2n}+t^{2n}} \label{eq:integrant}
\end{equation}
where
\begin{equation}
 t_k \equiv \left(\frac{k^2+m_0^2}{g_{2n}^2M_*^2}\right)^{1/2n}\frac{M_*}{v}.
\end{equation}
After analytic continuation, (\ref{eq:integrant}) has $2n$ poles at
\begin{equation}
 t = t_k e^{i\pi(m+1/2)/n}\equiv t_m
\end{equation}
for $m=0, 1, \cdots, 2n-1$.
At the neighborhood of $t\sim t_m$, one can obtain
\begin{eqnarray}
 \frac{\dot{\omega}_k}{2\omega_k}
  &=& \frac{1}{4}\frac{1}{t-t_m}\left(1+\mathcal{O}(t-t_m)\right) \\
 \int_{-\infty}^tdt'\omega_k(t')
  &=& \int_{-\infty}^{t_m}dt'\omega_k(t') \nonumber \\
  & & +\frac{2\sqrt{2n}g_{2n}v^nt_m^{n-1/2}}{3M_*^{n-1}}(t-t_m)^{3/2} \nonumber \\
  & & +\mathcal{O}\left((t-t_m)^{5/2}\right).
\end{eqnarray}
Choosing the integral path in the lower half, which corresponds to
choosing $m=n,n+1,\cdots, 2n-1$,
we find that eq.(\ref{eq:solution_of_beta}) can be rewritten as
\begin{equation}
 \beta_k(+\infty) \sim \sum_{m=n}^{2n-1} U_m \exp{\left[ -2i \int_{-\infty}^{t_m} dt' \omega_k(t') \right]}, \label{eq:beta2}
\end{equation}
where
\begin{equation}
  U_m \equiv \frac{1}{4} \int_{C_m} \frac{dt}{t-t_m}
   \exp{\left[-i\frac{4\sqrt{2n}g_{2n}v^nt_m^{n-1/2}}{3M_*^{n-1}}(t-t_m)^{3/2} \right]}. \label{eq:U}
\end{equation}
$C_m$ denotes the path approaching to and going around the pole at $t=t_m$
on the steepest descent path (and leaves the pole).  See Figure \ref{fig:path}.
%% figure %%%%%%%%%%%%%%%%%%%%%%%%%%%%%%%%%%%%%%%%%%%%%%%%%%%%%%%%%%%%%%%%%%%%%%%%%%%%%%%%%%%%%%%%%%%%%%%%%%%%%
%%%%%%%%%%%%%%%%%%%%%%%%%%%%%%%%%%%%%%%%%%%%%%%%%%%%%%%%%%%%%%%%%%%%%%%%%%%%%%%%%%%%%%%%%%%%%%%%%%%%%%%%%%%%%%%
If we take the angle going around the pole to be $4\pi/3$, the
outward-going path becomes the steepest descent path.
Then, we can obtain
\begin{equation}
 U_m \sim \frac{i\pi}{3}. \label{eq:U2}
\end{equation}
Using the above results, the distribution can be calculated as
\begin{eqnarray}
% n_k &=& |\beta_k|^2 \nonumber \\
%  &\sim& \frac{\pi^2}{9} \sum_{m,m'=n}^{2n-1}
%   \exp{\left[ -2i \left\{ \left( \int_{-\infty}^{\tilde{t}_m}dt \omega_k \right)
%   - \left( \int_{-\infty}^{\tilde{t}_{m'}}dt \omega_k \right)^* \right\} \right]} \nonumber \\
 |\beta_k|^2
   &\sim& \frac{\pi^2}{9} \sum_{m,m'=n}^{2n-1}
    \exp{\left[ -2i \left(\Omega_m-\Omega_{m'}^* \right) \right]} \label{eq:occupation_number}
\end{eqnarray}
where
\begin{eqnarray}
 \Omega_m
  &\equiv& \int_0^{t_m} dt' \omega_k(t') \\
  &=& \sqrt{k^2+m_0^2} \ t_k e^{i\pi(m+1/2)/n} \nonumber \\
  & & \quad \times \frac{\Gamma\left(\frac{3}{2}\right)\Gamma\left(1+\frac{1}{2n}\right)}{\Gamma\left(\frac{3}{2}+\frac{1}{2n}\right)}. \label{eq:Omega}
\end{eqnarray}

Especially, in the case of $n=1$, we can obtain
\begin{eqnarray}
 |\beta_k(+\infty)|^2
  &\sim& \frac{\pi^2}{9} e^{-2i \left(\Omega_1-\Omega_{1}^* \right)} \nonumber \\
  &=& \frac{\pi^2}{9} \exp\left[-\pi\frac{k^2+m_0^2}{g_2v}\right] \nonumber\\
  &=& \frac{\pi^2}{9} e^{-2K_{ud}}
\end{eqnarray}
where $K_{ud}$ is defined in (\ref{eq:K_ud}).
Hence, this result reproduces the results shown in \ref{sec:EWKB_E_pos_quardra} except for the prefactor
$\pi^2/9\sim 1.097.$
This extra factor seems to be due to the approximation by the steepest descent method.

For the case of $n=2$,  we can obtain
\begin{eqnarray}
 |\beta_k(+\infty)|^2
  &\sim& \frac{\pi^2}{9}\left(e^{-2i \left(\Omega_2-\Omega_2^* \right)}
   +e^{-2i \left(\Omega_3-\Omega_3^* \right)} \right. \nonumber \\
  & & \qquad \left. +e^{-2i \left(\Omega_2-\Omega_3^* \right)}
   +e^{-2i \left(\Omega_3-\Omega_2^* \right)} \right) \nonumber \\
  &=& \frac{\pi^2}{9} \cdot 2e^{-2.47\kappa_4^{3/2}}\left(1+\cos(2.47\kappa_4^{3/2})\right) \nonumber \\
  &=& \frac{\pi^2}{9} \cdot 2e^{-2K_{AB}}\left(1+\cos(2K_{AB})\right)
\end{eqnarray}
where $\kappa_4$ and $K_{AB}$ are defined in (\ref{eq:def_kappa_4}) and (\ref{eq:def_K_AB}), respectively.
This result also reproduces the analytic result by the EWKB method except for the prefactor.

\end{document}